\documentclass{elsart}
\usepackage{amsmath,amssymb,graphicx}
\journal{Physica A}

\newcommand{\uu}{\mbox{\boldmath$u$}}
\newcommand{\vv}{\mbox{\boldmath$v$}}
\newcommand{\xx}{\mbox{\boldmath$x$}}
\newcommand{\D}{{\mathrm{d}}}

\begin{document}

\begin{frontmatter}

\title{Nonequilibrium entropy limiters \\ in lattice Boltzmann methods\thanksref{titlefn}}

\author{R.~A.~Brownlee},
\ead{r.brownlee@mcs.le.ac.uk}
\author{A.~N.~Gorban\corauthref{cor}},
\corauth[cor]{Corresponding author.} \ead{a.gorban@mcs.le.ac.uk}
\author{J.~Levesley}
\ead{j.levesley@mcs.le.ac.uk}
\thanks[titlefn]{This work is supported by EPSRC grant number GR/S95572/01.}
\address{Department of Mathematics, University of Leicester, Leicester LE1 7RH, UK}

\begin{abstract}
We construct a system of nonequilibrium entropy limiters for the
lattice Boltzmann methods (LBM). These limiters erase spurious
oscillations without blurring of shocks, and do not affect smooth
solutions. In general, they do the same work for LBM as flux
limiters do for finite differences, finite volumes and finite
elements methods, but for LBM the main idea behind the construction
of nonequilibrium entropy limiter schemes is to transform a field of
a scalar quantity --- nonequilibrium entropy. There are two families
of limiters: (i) based on restriction of nonequilibrium entropy
(entropy ``trimming") and (ii) based on filtering of nonequilibrium
entropy (entropy filtering). The physical properties of LBM provide
some additional benefits: the control of entropy production and
accurate estimate of introduced artificial dissipation are possible.
The constructed limiters are tested on classical numerical examples:
1D athermal shock tubes with an initial density ratio 1:2 and the 2D
lid-driven cavity for Reynolds numbers $\mathrm{Re}$ between $2000$
and $7500$ on a coarse $100 \times 100$ grid. All limiter
constructions are applicable for both entropic and non-entropic
quasiequilibria.
\end{abstract}


\end{frontmatter}

\section{Introduction\label{Intro}}

In 1959, S.K. Godunov~\cite{Godunov} demonstrated that a (linear)
scheme for a PDE could not, at the same time, be monotone and second
order accurate. Hence, we should choose between spurious oscillation
in high order non-monotone schemes and additional dissipation in
first order schemes. Flux limiter schemes are invented to combine
high resolution schemes in areas with smooth fields and first order
schemes in areas with sharp gradients.

The idea of flux limiters can be illustrated by computation of the
flux $F_{0,1}$ of the conserved quantity $u$ between a cell marked
by 0 and one of two its neighbour cells marked by  $\pm 1$:
\begin{equation*}
   F_{0,1}=(1-\phi (r))f^{\rm low}_{0,1} + \phi (r) f^{\rm high}_{0,1},
\end{equation*}
where $f^{\rm low}_{0,\,  1}$, $f^{\rm high}_{0,\,  1}$ are low and
high resolution scheme fluxes, respectively,
$r=(u_0-u_{-1})/(u_1-u_0)$, and $\phi (r)\geq 0$ is a flux limiter
function. For $r$ close to 1, the flux limiter function $\phi (r)$
should be also close to 1.

Many flux limiter schemes have been invented during the last two
decades~\cite{Wess}. No particular limiter works well for all
problems, and a choice is usually made on a trial and error basis.

Below are several examples of flux limiter functions:
\begin{align*}
\phi_{mm} (r) &= \max \left[ 0 , \min \left( r , 1 \right) \right] \quad \text{(minmod,~\cite{Roe1986})}; \\
\phi_{os} (r) &= \max \left[ 0 , \min \left( r, \beta \right)\right],\; \left(1 \leq \beta \leq 2 \right) \quad \text{(Osher,~\cite{ChatOsher1983})}; \\
 \phi_{mc} (r) &= \max \left[ 0 , \min \left( 2 r, 0.5 (1+r), 2 \right) \right] \quad \text{(monotonised central~\cite{vanLeer1977})};\\
 \phi_{sb} (r) &= \max \left[ 0, \min \left( 2 r , 1 \right), \min \left( r, 2 \right) \right] \quad \text{(superbee,~\cite{Roe1986})}; \\
 \phi_{sw} (r) &= \max \left[ 0 , \min \left( \beta r, 1 \right),\left( r, \beta \right) \right],  \;   \left(1 \leq \beta \leq 2\right) \quad \text{(Sweby,~\cite{Sweby1984})}.
\end{align*}

The lattice Boltzmann method has been proposed as a discretization
of Boltzmann's kinetic equation and is now in wide use in fluid
dynamics and beyond (for an introduction and review
see~\cite{succi01}). Instead of fields of moments $M$, the lattice
Boltzmann method operates with fields of discrete distributions
$f$. This allows us to construct very simple limiters that do not
depend on slopes or gradients.

All the limiters we construct are based on the representation of
distributions $f$ in the form:
\begin{equation*}
f=f^*+\|f-f^*\| \frac{f-f^*}{\|f-f^*\|},
\end{equation*}
where $f^*$ is the correspondent quasiequilibrium (conditional
equilibrium) for given moments $M$, $f-f^*$ is the nonequilibrium
``part" of the distribution, which is represented in the form
``norm$\times$direction'' and $\|f-f^*\|$ is the norm of that
nonequilibrium component (usually this is the entropic norm).
Limiters change the norm of the nonequilibrium component $f-f^*$,
but do not touch its direction or the equilibrium. In particular,
limiters do not change the macroscopic variables, because moments
for $f$ and $f^*$ coincide. All limiters we use are transformations
of the form
\begin{equation}\label{LimGenF}
f \mapsto f^*+ \phi \times(f-f^*)
\end{equation}
with $\phi >0$. If $f-f^*$ is too big, then the limiter should
decrease its norm.

The outline of the paper is as follows. In Sec.~\ref{sec1} we
introduce the notions and notations from lattice Boltzmann theory
we need, in Sec.~\ref{sec2} we elaborate the idea of entropic
limiters in more detail and construct several nonequilibrium
entropy limiters for LBM, in Sec.~\ref{sec3} some numerical
experiments are described:
\begin{enumerate}
\item 1D athermal shock tube examples;
\item steady state vortex centre locations and observation of first Hopf
bifurcation in 2D lid-driven cavity flow.
\end{enumerate}
Concluding remarks are given in Sec.~\ref{sec4}.

\section{Background}\label{sec1}

The essence of lattice Boltzmann methods was formulated by S.~Succi
in the following maxim: ``Nonlinearity is local, non-locality is
linear''\footnote{S. Succi, ``Lattice Boltzmann at all-scales: from
turbulence to DNA translocation'', Mathematical Modelling Centre
Distinguished Lecture, University of Leicester, Leicester UK,
15\textsuperscript{th} November 2006.}. We should even strengthen
this statement. Non-locality (a) is linear; (b) is exactly and
explicitly solvable for all time steps; (c) space discretization is
an exact operation.

The lattice Boltzmann method is a discrete velocity method. The
finite set of velocity vectors $\{\vv_i\}$ ($i=1,...m$) is selected,
and a fluid is described by associating, with each velocity $\vv_i$,
a single-particle distribution function $f_i=f_i(\xx,t)$ which is
evolved by advection and interaction (collision) on a fixed
computational lattice. The values $f_i$ are named
\emph{populations}. If we look at all lattice Boltzmann models, one
finds that there are two steps: free flight for time $\delta t$ and
a local collision operation.

The free flight transformation for continuous space is
\begin{equation*}
f_i(\xx,t+\delta t) = f_i(\xx-\vv_i \delta t ,t).
\end{equation*}
After the free flight step the collision step follows:
\begin{equation}\label{coll}
f_i(\xx) \mapsto F_i (\{f_j(\xx)\}),
\end{equation}
or in the vector form
\begin{equation*}
f(\xx) \mapsto F (f(\xx)).
\end{equation*}
Here, the  \emph{collision operator} $F$ is the set of functions
$F_i (\{f_j\})$ ($i=1,...m$). Each function $F_i$ depends on all
$f_j$ ($j=1,...m$): new values of the populations $f_i$ at a point
$\xx$ are known functions of all previous population values at the
same point.

The lattice Boltzmann chain ``free flight $\to$ collision $\to$ free
flight $\to$ collision $\dotsb$'' can be exactly restricted onto any
space lattice which is invariant with respect to space shifts of the
vectors $\vv_i \delta t$ ($i=1,...m$). Indeed, free flight
transforms the population values at sites of the lattice into the
population values at sites of the same lattice. The collision
operator~\eqref{coll} acts pointwise at each lattice site
separately. Much effort has been applied to answer the questions:
``how does the lattice Boltzmann chain approximate the transport
equation for the moments $M$?'', and ``how does one construct the
lattice Boltzmann model for a given macroscopic transport
phenomenon?'' (a review is presented in book  \cite{succi01}).

In our paper we propose a universal construction of limiters for all
possible collision operators, and the detailed construction of
$F_i(\{f_j\})$ is not important for this purpose. The only part of
this construction we use is the local equilibria (sometimes these
states are named conditional equilibria, quasiequilibria, or even
simpler, equilibria).

The lattice Boltzmann models should describe the macroscopic
dynamic, i.e., the dynamic of macroscopic variables. The macroscopic
variables $M_\ell(\xx)$ are some linear functions of the population
values at the same point: $M_\ell (\xx)=\sum_i m_{\ell i} f_i(\xx)$,
or in the vector form, $M(\xx)=m(f(\xx))$. The macroscopic variables
are invariants of collisions:
\begin{equation*}
    \sum_i m_{\ell i} f_i=\sum_i m_{\ell i} F_i (\{f_j\}) \quad \text{(or $m(f)=m(F(f))$).}
\end{equation*}
The standard example of the macroscopic variables are hydrodynamic
fields (density--velocity--energy density): $\{n, n \uu,
E\}(\xx):=\sum_i \{1, \vv_i,  v_i^2/2\} f_i(\xx)$. But this is not
an obligatory choice. If we would like to solve, by LBM methods, the
Grad equations~\cite{Grad} or some extended thermodynamic
equations~\cite{EIT}, we should extend the list of moments (but, at
the same time, we should be ready to introduce more discrete
velocities for a proper description of these extended moment
systems). On the other hand, the athermal lattice Boltzmann models
with a shortened list of macroscopic variables $\{n, n \uu\}$ are
very popular.

The quasiequilibrium is the positive fixed point of the collision
operator for the given macroscopic variables $M$. We assume that
this point exists, is unique and depends smoothly on $M$. For the
quasiequilibrium population vector for given $M$ we use the notation
$f^*_M$, or simply $f^*$, if the correspondent value of $M$ is
obvious. We use $\Pi^*$ to denote the equilibration projection
operation of a distribution $f$ into the corresponding
quasiequilibrium state:
\begin{equation*}
\Pi^*(f)=f^*_{m(f)}.
\end{equation*}
For some of the collision models an entropic description of
equilibrium is possible: an entropy density function $S(f)$ is
defined and the quasiequilibrium point $f^*_M$ is the entropy
maximiser for given $M$~\cite{KGSBPRL,LB3}.

As a basic example we shall consider the lattice
Bhatnagar--Gross--Krook (LBGK) model with overrelaxation (see,
e.g.,~\cite{Benzi,LB1,Higuera,karlin06,succi01}). The LBGK collision
operator is
\begin{equation}\label{collLBGK}
F(f) :=\Pi^*(f) + (2\beta -1)(\Pi^*(f)-f),
\end{equation}
where $\beta \in [0,1]$. For $\beta = 0$, LBGK collisions do not
change $f$, for $\beta = 1/2$ these collisions act as equilibration
(this corresponds to the Ehrenfests' coarse graining
\cite{ehrenfest11} further developed in
\cite{Raz,GKOeTPRE2001,gorban06}), for $\beta = 1$, LBGK collisions
act as a point reflection with the center at the quasiequilibrium
$\Pi^*(f)$.

It is shown~\cite{BGJ} that under some stability conditions and
after an initial period of relaxation, the simplest LBGK collision
with overrelaxation~\cite{Higuera,succi01} provides second order
accurate approximation for the macroscopic transport equation with
viscosity proportional to $\delta t (1-\beta)/ \beta$.

The entropic LBGK (ELBM)
method~\cite{boghosian01,gorban06,KGSBPRL,LB3} differs in the
definition of~\eqref{collLBGK}: for $\beta=1$ it should conserve the
entropy, and in general has the following form:
\begin{equation}\label{elbm}
F(f) := (1-\beta)f+\beta \tilde{f},
\end{equation}
where $\tilde{f}= (1-\alpha) f+\alpha \Pi^*(f)$. The number
$\alpha=\alpha(f)$ is chosen so that the constant entropy condition
is satisfied: $S(f)=S(\tilde{f})$. For LBGK~\eqref{collLBGK},
$\alpha=2$. Of course, for ELBM the entropic definition of
quasiequilibrium should be valid.

In the low-viscosity regime, LBGK suffers from numerical
instabilities which readily manifest themselves as local blow-ups
and spurious oscillations.

The LBM experiences the same spurious oscillation problems near
sharp gradients as high order schemes do. The physical properties
of the LBM schemes allows one to construct new types of limiters:
the nonequilibrium entropy limiters. In general, they do the same
work for LBM as flux limiters do for finite differences, finite
volumes and finite elements methods, but for LBM the main idea
behind the construction of nonequilibrium entropy limiter schemes
is to limit a scalar quantity
--- nonequilibrium entropy (and not the vectors or tensors of spatial
derivatives, as it is for flux limiters). These limiters introduce
some additional dissipation, but all this dissipation could easily
be evaluated through analysis of nonequilibrium entropy
production.

Two examples of such limiters have been recently proposed: the
positivity rule~\cite{Rob_preprint,li04,Tosi06} and the Ehrenfests'
regularisation \cite{Rob}. The positivity rule just provides
positivity of distributions: if a collision step produces negative
populations, then the positivity rule returns them to the boundary
of positivity. In the Ehrenfests' regularisation, one selects the
$k$ sites with highest nonequilibrium entropy (the difference
between entropy of the state $f$ and entropy of the corresponding
quasiequilibrium state $f^*$ at a given space point) that exceed a
given threshold and equilibrates the state in these sites.

The positivity rule and Ehrenfests' regularisation provide rare,
intense and localised corrections. It is easy and also
computationally cheap to organise more gentle transformation with
smooth shift of highly nonequilibrium states to quasiequilibrium.
The following regularisation transformation distributes its action
smoothly: we can just choose in (\ref{LimGenF}) $\phi=\phi (\Delta
S(f))$ with sufficiently smooth function $\phi (\Delta S(f))$. Here
$f$ is the state at some site, $f^*$ is the corresponding
quasiequilibrium state, $S$ is entropy, and $\Delta
S(f):=S(f^*)-S(f)$.

The next step in the development of the nonequilibrium entropy
limiters is in the usage of local entropy filters. The filter of
choice here is the median filter: it does not erase sharp fronts,
and is much more robust than convolution filters.

An important problem is: ``how does one create nonequilibrium
entropy limiters for LBM with non-entropic quasiequilibria?''. We
propose a solution of this problem based on the nonequilibrium
Kullback entropy. For entropic quasiequilibrium the Kullback entropy
approach gives the same entropic limiters. In thermodynamics,
Kullback entropy belongs to the family of Massieu--Planck--Kramers
functions (canonical or grandcanonical potentials).

\section{Nonequilibrium entropy limiters for LBM \label{sec2}}

\subsection{Positivity rule} 
\begin{figure}
\begin{centering}
\includegraphics[width=7cm]{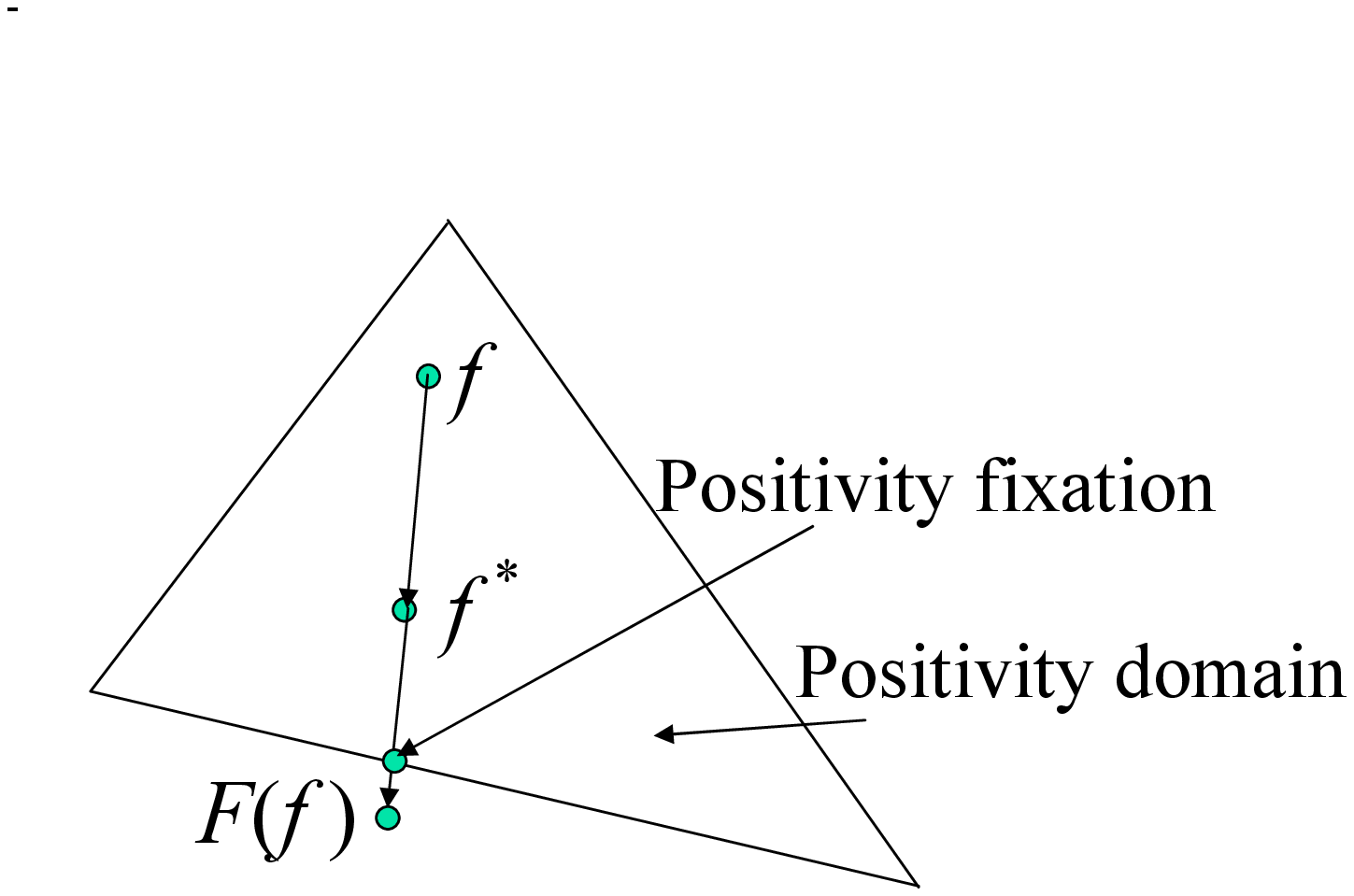}
\caption{\label{PosRule}Positivity rule in action. The motions
stops at the positivity boundary.}
\end{centering}
\end{figure}

There is a simple recipe for positivity
preservation~\cite{Rob_preprint,li04,Tosi06}: to substitute
nonpositive $I_0^{\beta}(f)(\xx)$ by the closest nonnegative state
that belongs to the straight line
\begin{equation}\label{StrLine}
  \Bigr\{\lambda f(\xx) + (1-\lambda) \Pi^*(f(\xx))|\: \lambda \in
\mathbb{R}\Bigl\}
\end{equation}
defined by the two points, $f(\xx)$ and corresponding
quasiequilibrium. This operation is to be applied pointwise, at
points of the lattice where positivity is violated. The coefficient
$\lambda$ depends on $\xx$ too. Let us call this recipe the
\textit{positivity rule} (Fig.~\ref{PosRule}). This recipe preserves
positivity of populations and probabilities, but can affect accuracy
of approximation. The same rule is necessary for ELBM~\eqref{elbm}
when the positive ``mirror state'' $\tilde{f}$ with the same entropy
as $f$ does not exists on the straight line~\eqref{StrLine}.

\subsection{Ehrenfests' regularisation}

To discuss methods with additional dissipation, the entropic
approach is very convenient. Let entropy $S(f)$ be defined for each
population vector $f=(f_i)$ (below we use the same letter $S$ for
local in space entropy, and hope that context will make this
notation always clear). We assume that the global entropy is a sum
of local entropies for all sites. The local nonequilibrium entropy
is
\begin{equation}\label{locaNEQentropy}
 \Delta S(f):= S(f^*) - S(f),
\end{equation}
where $f^*$ is the corresponding local quasiequilibrium at the same
point.

The \emph{Ehrenfests' regularisation}~\cite{Rob_preprint,Rob}
provides ``entropy trimming'': we monitor local deviation of $f$
from the corresponding quasiequilibrium, and when $\Delta S(f)(\xx)$
exceeds a pre-specified threshold value $\delta$, perform local
Ehrenfests' steps to the corresponding quasiequilibrium: $f \mapsto
f^*$ at those points.

So that the Ehrenfests' steps are not allowed to degrade the
accuracy of LBGK it is pertinent to select the $k$ sites with
highest $\Delta S>\delta$. The a posteriori estimates of added
dissipation could easily be performed by analysis of entropy
production in Ehrenfests' steps. Numerical experiments show (see,
e.g.,~\cite{Rob_preprint,Rob}) that even a small number of such
steps drastically improve stability.

To avoid the change of accuracy order ``on average'', the number of
sites with this step should be $\leq \mathcal{O}(Nh/L)$ where $N$ is
the total number of sites, $h$ is the step of the space
discretization and $L$ is the macroscopic characteristic length. But
this rough estimate of accuracy in average might be destroyed by
concentration of Ehrenfests' steps in the most nonequilibrium areas,
for example, in the boundary layer. In that case, instead of the
total number of sites $N$ in $\mathcal{O}(Nh/L)$ we should take the
number of sites in a specific region. The effects of concentration
could be easily analysed a posteriori.

\subsection{Smooth limiters of nonequilibrium entropy}

The positivity rule and Ehrenfests' regularisation provide rare,
intense and localised corrections. Of course, it is easy and also
computationally cheap to organise more gentle transformation with a
smooth shift of highly nonequilibrium states to quasiequilibrium.
The following regularisation transformation distributes its action
smoothly:
\begin{equation}\label{smoothReg}
f \mapsto f^*+ \phi (\Delta S(f))(f-f^*).
\end{equation}
The choice of function $\phi$ is highly ambiguous, for example,
$\phi=1/(1+ \alpha \Delta S^k)$ for some $\alpha>0$ and $k>0$. There
are two significantly different choices: (i) ensemble-independent
$\phi$ (i.e., the value of $\phi$ depends on local value of $\Delta
S$ only) and (ii) ensemble-dependent $\phi$, for example
\begin{equation}\label{phiExamp}
\phi(\Delta S)=\frac{1+(\Delta S/(\alpha {\rm E} (\Delta
S)))^{k-1/2}} {1+(\Delta S/(\alpha {\rm E} (\Delta S)))^{k}},
\end{equation}
where ${\rm E}( \Delta S)$ is the average value of $\Delta S$ in the
computational area, $k\geq 1$, and $\alpha \gtrsim 1$. For small
$\Delta S$, $\phi(\Delta S) \approx 1$ and for $\Delta S \gg \alpha
{\rm E} (\Delta S)$, $\phi(\Delta S)$ tends to $\sqrt{\alpha {\rm E}
(\Delta S) / \Delta S}$. It is easy to select an ensemble-dependent
$\phi$ with control of total additional dissipation.

\subsection{Monitoring of total dissipation}

For given $\beta$, the entropy production in one LBGK step in
quadratic approximation for $\Delta S$ is:
\begin{equation*}
\delta_{\rm LBGK} S \approx [1-(2\beta-1)^2] \sum_{x} \Delta S
(\xx),
\end{equation*}
where $\xx$ is the grid point, $\Delta S (\xx)$ is nonequilibrium
entropy~\eqref{locaNEQentropy} at point $\xx$, $\delta_{\rm LBGK} S$
is the total entropy production in a single LBGK step. It would be
desirable if the total entropy production for the limiter
$\delta_{\rm lim} S$ was small relative to $\delta_{\rm LBGK} S$:
\begin{equation}\label{LimEntLim}
\delta_{\rm lim} S  < \delta_0 \delta_{\rm LBGK} S.
\end{equation}

A simple ensemble-dependent limiter (perhaps, the simplest one) for
a given $\delta_0$ operates as follows. Let us collect the histogram
of the $\Delta S(\xx)$ distribution, and estimate the distribution
density, $p(\Delta S)$. We have to estimate a value $\Delta S_0$
that satisfies the following equation:
\begin{equation}\label{EntThresh}
\int_{\Delta S_0}^{\infty} p(\Delta S) (\Delta S - \Delta S_0) \,
\D \Delta S = \delta_0 [1-(2\beta-1)^2] \int_0^{\infty} p(\Delta
S) \Delta S \, \D \Delta S.
\end{equation}
In order not to affect distributions with small expectation of
$\Delta S$, we choose a threshold $\Delta S_{\rm t}= \max \{ \Delta
S_0, \delta\}$, where $\delta$ is some predefined value (as in the
Ehrenfests' regularisation). For states at sites with $\Delta S \geq
\Delta S_{\rm t} $ we provide homothety with quasiequilibrium center
 $f^*$ and coefficient $\sqrt{\Delta S_{\rm t} /
\Delta S}$ (in quadratic approximation  for nonequilibrium entropy):
\begin{equation}\label{entrHom}
f(\xx) \mapsto f^*(\xx) + \sqrt{\frac{\Delta S_{\rm t}}{\Delta S}}
(f(\xx)- f^*(\xx)) .
\end{equation}

\subsection{Median entropy filter}

The limiters described above provide pointwise correction of
nonequilibrium entropy at the ``most nonequilibrium'' points. Due to
the pointwise nature, the technique does not introduce any
nonisotropic effects, and provides some other benefits. But if we
involve the local structure, we can correct local non-monotone
irregularities without touching regular fragments. For example, we
can discuss monotone increase or decrease of nonequilibrium entropy
as regular fragments and concentrate our efforts on reduction of
``speckle noise'' or ``salt and pepper noise''. This approach allows
us to use the accessible resource of entropy
change~\eqref{LimEntLim} more thriftily.

Among all possible filters, we suggest the~\emph{median filter}. The
median is a more robust average than the mean (or the weighted mean)
and so a single very unrepresentative value in a neighborhood will
not affect the median value significantly. Hence, we suppose that
the median entropy filter will work better than entropy convolution
filters.

The median filter considers each site in turn and looks at its
nearby neighbours. It replaces the nonequilibrium entropy value
$\Delta S$ at the point with the median of those values $\Delta
S_{\rm med}$, then updates $f$ by the transformation~\eqref{entrHom}
with the homothety coefficient $\sqrt{\Delta S_{\rm med}/{\Delta
S}}$. The median, $\Delta S_{\rm med}$, is calculated by first
sorting all the values from the surrounding neighbourhood into
numerical order and then replacing that being considered with the
middle value. For example, if a point has 3 nearest neighbors
including itself, then after sorting we have 3 values $\Delta S$:
$\Delta S_1 \leq \Delta S_2 \leq  \Delta S_3$. The median value is
$\Delta S_{\rm med}=\Delta S_2$. For 9 nearest neighbors (including
itself) we have after sorting $\Delta S_{\rm med}=\Delta S_5$. For
27 nearest neighbors  $\Delta S_{\rm med}=\Delta S_{14}$.

We accept only dissipative corrections (those resulting in a
decrease of $\Delta S$, $\Delta S_{\rm med}< \Delta S$) because of
the second law of thermodynamics. The analogue of~\eqref{EntThresh}
is also useful for acceptance of the most significant corrections.

Median filtering is a common step in image processing~\cite{Pratt}
for the smoothing of signals and the suppression of impulse noise
with preservation of edges.

\subsection{Entropic steps for non-entropic quasiequilibria}

Beyond the quadratic approximation for nonequilibrium entropy all
the logic of the above mentioned constructions remain the same.
There exists only one significant change: instead of a simple
homothety~\eqref{entrHom} with coefficient $\sqrt{\Delta S_{\rm
t}/{\Delta S}}$ the transformation~\eqref{smoothReg} should be
applied, where the multiplier $\phi$ is a solution of the nonlinear
equation
\begin{equation*}
    S(f^*+ \phi (f-f^*))= S(f^*)-\Delta S_{\rm t}.
\end{equation*}
This is essentially the same equation that appears in the definition
of ELBM steps~\eqref{elbm}.

More differences emerge for LBM with non-entropic quasiequilibria.
The main idea here is to reason that non-entropic quasiequilibria
appear only because of technical reasons, and approximate continuous
physical entropic quasiequilibria. This is not an approximation of a
density function, but an approximation of measure, i.e., from the
cubature formula:
\begin{align*}
f(\vv) &\approx \sum_i f_i  \delta (\vv -\vv _i)  \\
 \int \varphi (\vv) f(\vv) \, \D \vv &\approx \sum_i \varphi (\vv_i)
 f_i.
\end{align*}
The discrete populations $f_i$ are connected to continuous (and
sufficiently smooth) densities $f(\vv)$ by cubature weights $f_i
\approx w_i f(\vv_i)$. These weights for quasiequilibria are found
by moment and flux matching conditions~\cite{HudongGrad}. It is
impossible to approximate the BGS entropy $\int f \ln f \D \vv$ just
by discretization (to change integration by summation, and
continuous distribution $f$ by discrete $f_i$), because cubature
weights appear as additional variables. Nevertheless, the
approximate discretization of the Kullback entropy $S_{\rm
K}$~\cite{Kull} does not change its form:
\begin{equation}\label{Kul}
S_{\rm K}(f)=- \int f(\vv) \ln\left(\frac{f(\vv)}{f^*(\vv)}\right)
\, \D \vv \approx - \sum_i f_i \ln\left(\frac{f_i}{f_i^*}\right),
\end{equation}
because ${f_i}/{f_i^*}$ approximates the ratio of functions
${f(\vv)}/{f^*(\vv)}$ and $\sum_i f_i\dotsc$ gives the integral
$\int f(\vv)\dotsc\D \vv$ approximation. Here, in~\eqref{Kul}, the
state $f^*$ is the quasiequilibrium with the same values of the
macroscopic variables as $f$. Moreover, for given values of the
macroscopic variables, $S_{\rm K}(f)$ achieves its maximum at the
point $f=f^*$ (both for continuous and for discrete distributions).
The corresponding maximal value is zero. Below, $S_{\rm K}$ is the
discrete Kullback entropy. If the approximate discrete
quasiequilibrium $f^*$ is non-entropic, we can use $-S_{\rm K}(f)$
instead of $ \Delta S(f)$.

For entropic quasiequilibria with perfect entropy the discrete
Kullback entropy gives the same $\Delta S$: $-S_{\rm K}(f)= \Delta
S(f)$. Let the discrete entropy have the standard form for an ideal
(perfect) mixture~\cite{karlin99}.
\begin{equation*}
S(f)=-\sum_i f_i \ln\biggl(\frac{f_i}{W_i}\biggr).
\end{equation*}
After the classical work of Zeldovich~\cite{Zeld}, this function is
recognised as a useful instrument for the analysis of kinetic
equations (especially in chemical kinetics~\cite{Kagan}). If we
define $f^*$ as the conditional entropy maximum for given
$M_j=\sum_k m_{jk} f_k$, then
\begin{equation*}
\ln f^*_k=\sum_j \mu_j m_{jk},
\end{equation*}
where $\mu_j(M)$ are the Lagrange multipliers (or ``potentials'').
For this entropy and conditional equilibrium we find
\begin{equation}\label{DelS}
\Delta S= S(f^*)-S(f)=\sum_i f_i \ln\biggl(\frac{f_i}{f^*_i}\biggr),
\end{equation}
if $f$ and $f^*$ have the same moments, $m(f)=m(f^*)$. The right
hand side of~\eqref{DelS} is $-S_{\rm K}(f)$.

In thermodynamics, the Kullback entropy belongs to the family of
Massieu--Planck--Kramers functions (canonical or grandcanonical
potentials). There is another sense of this quantity: $S_K$ is the
relative entropy of $f$ with respect to $f^*$~\cite{G1,Qian}.

In quadratic approximation,
\begin{equation*}
-S_{\rm K}(f)=\sum_i f_i \ln\biggl(\frac{f_i}{f^*_i}\biggr) \approx
\sum_i\frac{(f_i-f^*_i)^2}{f^*_i}.
\end{equation*}

\subsection{ELBM collisions as a smooth limiter}

On the base of numerical tests, the authors of~\cite{Tosi06} claim
that the positivity rule provides the same results (in the sense of
stability and absence/presence of spurious oscillations) as the ELBM
models, but ELBM provides better accuracy.

For the formal definition of ELBM~\eqref{elbm} our tests do not
support claims that ELBM erases spurious oscillations (see below).
Similar observation for Burgers equation was previously published
in~\cite{Bruce}. We understand this situation in the following way.
The entropic method consists at least of three components:
\begin{enumerate}
\item{entropic quasiequilibrium, defined by entropy
maximisation;} \item{entropy balanced collisions~\eqref{elbm} that
have to provide proper entropy balance;} \item{a method for the
solution of the transcendental equation $S(f)=S(\tilde{f})$ to find
$\alpha=\alpha(f)$ in~\eqref{elbm}.}
\end{enumerate}
It appears that the first two items do not affect spurious
oscillations at all, if we solve the equation for $\alpha(f)$ with
high accuracy. Additional viscosity is, potentially, added by
explicit analytic formulas for $\alpha(f)$. In order not to decrease
entropy, errors in these formulas always increase dissipation. This
can be interpreted as a hidden transformation of the
form~\eqref{smoothReg}, where the coefficients in $\phi$ depend also
on $f^*$.

\subsection{Monotonic and double monotonic limiters}

Two monotonicity properties are important in the theory of
nonequilibrium entropy limiters:
\begin{enumerate}
\item a limiter should move the distribution to equilibrium: in all
cases of~\eqref{LimGenF} $0 \leq \phi \leq 1$. This is the \emph{
dissipativity} condition which means that limiters never produce
negative entropy.
\item a limiter should not change the order of states on the line:
if for two distributions with the same moments, $f$ and $f'$,
$\Delta S (f) > \Delta S (f')$ before the limiter transformation,
then the same inequality should hold after the limiter
transformation too. For example, for the limiter~\eqref{smoothReg}
it means that $\Delta S(f^*+x \phi (\Delta S(f^*+x(f-f^*))(f-f^*))$
is a monotonically increasing function of $x>0$.
\end{enumerate}
In quadratic approximation,
\begin{align*}
\Delta S(f^*+x(f-f^*)) &= x^2 \Delta S(f),\\
\Delta S(f^*+x \phi (\Delta S(f^*+x(f-f^*))(f-f^*)) &= x^2
\phi^2(x^2 \Delta S(f)),
\end{align*}
and the second monotonicity condition transforms into the following
requirement: $y \phi (y^2 s)$ is a monotonically increasing (not
decreasing) function of $y>0$ for any $s>0$.

If a limiter satisfies both monotonicity conditions, we call it
``double monotonic''. For example, Ehrenfests' regularisation
satisfies the first monotonicity condition, but obviously violates
the second one. The limiter~\eqref{phiExamp} violates the first
condition for small $\Delta S$, but is dissipative and satisfies the
second one in quadratic approximation for large $\Delta S$. The
limiter with $\phi=1/(1+ \alpha \Delta S^k)$ always satisfies the
first monotonicity condition, violates the second if $k > 1/2$, and
is double monotonic (in quadratic approximation for the second
condition), if $0 < k \leq 1/2$. The threshold
limiters~\eqref{entrHom} are also double monotonic. Of course, it is
not forbidden to use any type of limiters under the local and global
control of dissipation, but double monotonic limiters provide some
natural properties automatically, without additional care.

\section{Numerical experiment}\label{sec3}

To conclude this paper we report some numerical experiments
conducted to demonstrate the performance of some of the proposed
nonequilibrium entropy limiters for LBM from Sec.~\ref{sec2}.

\subsection{Velocities and quasiequilibria}

We will perform simulations using both entropic and non-entropic
quasiequilibria, but we always work with an athermal LBM model.
Whenever we use non-entropic quasiequilibria we employ Kullback
entropy~\eqref{DelS}.

In 1D, we use a lattice with spacing and time step $\delta t=1$ and
a discrete velocity set $\{v_1,v_2,v_3\}:=\{0,-1,1\}$ so that the
model consists of static, left- and right-moving populations only.
The subscript $i$ denotes population (not lattice site number) and
$f_1$, $f_2$ and $f_3$ denote the static, left- and right-moving
populations, respectively. The entropy is $S=-H$, with
\begin{equation*}
 H = f_1 \log(f_1/4)+f_2\log(f_2)+f_3 \log(f_3),
\end{equation*}
(see, e.g.,~\cite{karlin99}) and, for this entropy, the local
entropic quasiequilibrium state $f^*$ is available explicitly:
\begin{equation}\label{entropiceq1d}
\begin{split}
  f_1^{*} &= \frac{2 \rho}{3} \bigl(2-\sqrt{1+3 u^2}\bigr), \\
  f_2^{*} &= \frac{\rho}{6} \bigl((3u-1)+2\sqrt{1+3 u^2} \bigr),\\
  f_3^{*} &= -\frac{\rho}{6} \bigl((3u+1)-2\sqrt{1+3 u^2} \bigr),
\end{split}
\end{equation}
where
\begin{equation}\label{macro_dis}
  \rho := \sum_i f_i,\qquad u:=\frac{1}{ \rho} \sum_i v_i f_i.
\end{equation}
The standard non-entropic polynomial quasiequilibria~\cite{succi01}
are:
\begin{equation}\label{polyeq1d}
\begin{split}
    f_1^* &= \frac{2\rho}{3}\biggl(1-\frac{3u^2}{2}\biggr),\\
    f_2^* &= \frac{\rho}{6}(1-3u+3u^2),\\
    f_3^* &= \frac{\rho}{6}(1+3u+3u^2).
\end{split}
\end{equation}

In 2D, we employ a uniform $9$-speed square lattice with discrete
velocities $\{ \vv_i \, | \, i=0,1,\ldots 8\}$: $\vv_0=0$, $\vv_i =
(\cos((i-1)\pi/{2}),\sin((i-1){\pi}/{2}))$  for $i=1,2,3,4$, $\vv_i
= \sqrt{2} (\cos((i-5)\frac{\pi}{2}+\frac{\pi}{4}),
\sin((i-5)\frac{\pi}{2}+\frac{\pi}{4}))$ for $i=5,6,7,8$. The
numbering $f_0$, $f_1, \dotsc, f_8$  are for the static, east,
north, west, south, northeast, northwest, southwest and
southeast-moving populations, respectively. As usual, the entropic
quasiequilibrium state, $f^*$, can be uniquely determined by
maximising an entropy functional
\begin{equation*}
S(f) = -\sum_i f_i \log{\Bigl(\frac{f_i}{W_i}\Bigr)},
\end{equation*}
subject to the constraints of conservation of mass and
momentum~\cite{ansumali03}:
\begin{equation}\label{maxwellian}
  f_i^* = \rho W_i \prod_{j=1}^2
  \Bigl(2-\sqrt{1+3u_j^2}\Bigr)\Biggl( \frac{2u_j+\sqrt{1+3u_j^2}}{1-u_j}
  \Biggr)^{v_{i,j}}.
\end{equation}
Here, the \textit{lattice weights,} $W_i$, are given
lattice-specific constants: $W_0=4/9$, $W_{1,2,3,4}=1/9$ and
$W_{5,6,7,8}=1/36$. Analogously to~\eqref{macro_dis}, the
macroscopic variables $\rho$ and $\uu=(u_1,u_2)$ are the zeroth and
first moments of the distribution $f$, respectively. The standard
non-entropic polynomial quasiequilibria~\cite{succi01} are:
\begin{equation}\label{polyeq2d}
    f_i^* = \rho W_i \biggl(1+3 \vv_i\uu
    +\frac{9(\vv_i\uu)^2}{2}-\frac{3\uu^2}{2}\biggr).
\end{equation}

\subsection{LBGK and ELBM}
The governing equations for LBGK are
\begin{equation}\label{lbgkLB}
    f_i(x+v_i,t+1) =   f_i^*(x,t)+(2\beta-1)(f_i^*(x,t)-\!f_i(x,t)),
\end{equation}
where $\beta=1/(2\nu+1)$.

For ELBM~\eqref{elbm} the governing equations are:
\begin{equation}\label{elbmLB}
    f_i(x+v_i,t+1) = (1-\beta)f_i^*(x,t)+\beta \tilde{f}_i(x,t),
\end{equation}
with $\beta$ as above and $\tilde{f}=(1-\alpha)f+\alpha f^*$. The
parameter, $\alpha$, is chosen to satisfy a constant entropy
condition. This involves finding the nontrivial root of the equation
\begin{equation}\label{entropy_estimate}
  S((1-\alpha) f+\alpha f^*) = S(f).
\end{equation}
To solve~\eqref{entropy_estimate} numerically we employ a robust
routine based on bisection. The root is solved to an accuracy of
$10^{-15}$ and we always ensure that the returned value of $\alpha$
does not lead to a numerical entropy decrease. We stipulate that if,
at some site, no nontrivial root of~\eqref{entropy_estimate} exists
we will employ the positivity rule instead (Fig.~\ref{PosRule}).

\subsection{Shock tube}

The $1$D shock tube for a compressible athermal fluid is a standard
benchmark test for hydrodynamic codes.  Our computational domain
will be the interval $[0,1]$ and we discretize this interval with
$801$ uniformly spaced lattice sites. We choose the initial density
ratio as 1:2 so that for $x\leq 400$ we set $\rho=1.0$ else we set
$\rho=0.5$. We will fix the kinematic viscosity of the fluid at
$\nu=10^{-9}$.

\subsubsection{Comparison of LBGK and ELBM}

\begin{figure}
\begin{centering}
\includegraphics[width=13.0cm]{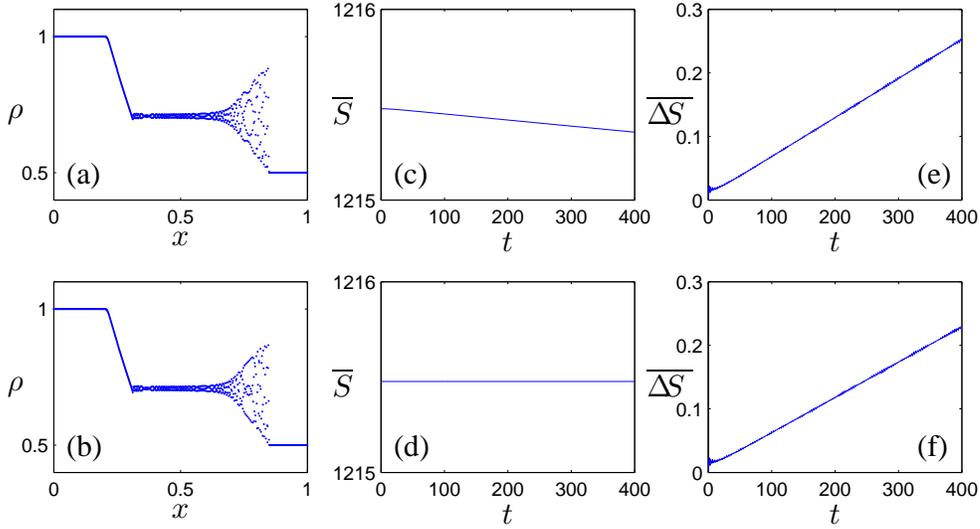}
\caption{Density and profile of the 1:2 athermal shock tube
simulation with $\nu=10^{-9}$ after $400$ time steps using (a)
LBGK~\eqref{lbgkLB}; (b) ELBM~\eqref{elbmLB}. In this example, no
negative population are produced by any of the methods so the
positivity rule is redundant. For ELBM in this
example,~\eqref{entropy_estimate} always has a nontrivial root.
Total entropy and nonequilibrium entropy time histories are shown in
panels (c), (d) and (e), (f) for LBGK and ELBM,
respectively.\label{lbgk_elbm}}
\end{centering}
\end{figure}

In Fig.~\ref{lbgk_elbm} we compare the shock tube density profile
obtained with LBGK (using entropic
quasiequilibria~\eqref{entropiceq1d}) and ELBM. On the same panel we
also display both the total entropy $\overline{S}(t) := \sum_x
S(x,t)$ and total nonequilibrium entropy $\overline{\Delta
S}(t):=\sum_x \Delta S(x,t)$ time histories. As expected, by
construction, we observe that total entropy is (effectively)
constant for ELBM. On the other hand, LBGK behaves non-entropically
for this problem. In both cases we observe that nonequilibrium
entropy grows with time.

\begin{figure}
\begin{centering}
\includegraphics[width=13cm]{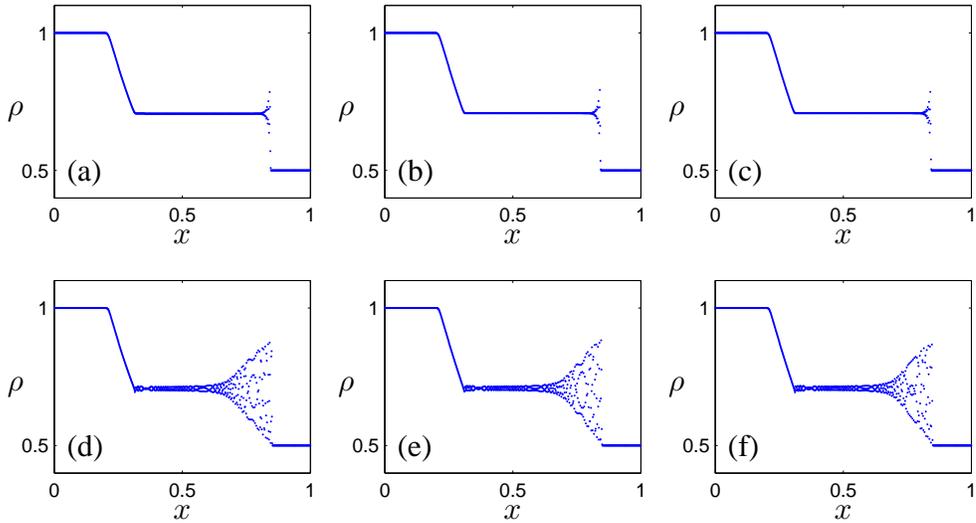}\\
\caption{Density and velocity profile of the 1:2 isothermal shock
tube simulation after $400$ time steps using (a) LBGK~\eqref{lbgkLB}
with polynomial quasiequilibria~\eqref{polyeq1d} [$\nu=3.3333\times
10^{-2}$]; (b) LBGK~\eqref{lbgkLB} with entropic
quasiequilibria~\eqref{entropiceq1d} [$\nu=3.3333\times 10^{-2}$];
(c) ELBM~\eqref{elbmLB} [$\nu=3.3333\times 10^{-2}$]; (d)
LBGK~\eqref{lbgkLB} with polynomial quasiequilibria~\eqref{polyeq1d}
[$\nu=10^{-9}$]; (e) LBGK~\eqref{lbgkLB} with entropic
quasiequilibria~\eqref{entropiceq1d} [$\nu=10^{-9}$]; (f)
ELBM~\eqref{elbmLB} [$\nu=10^{-9}$].\label{poly}}
\end{centering}
\end{figure}

As we can see, the choice between the two collision formulas
LBGK~\eqref{lbgkLB} or ELBM~\eqref{elbmLB} does not affect spurious
oscillation, and reported regularisation~\cite{karlin07} is,
perhaps, the result of approximate analytical solution of the
equation~\eqref{entropy_estimate}. Inaccuracy in the solution
of~\eqref{entropy_estimate} can be interpreted as a hidden
nonequilibrium entropy limiter. But it should be mentioned that the
entropic method consists not only of the collision formula, but,
what is important, includes a special choice of quasiequilibrium
that could improve stability (see, e.g.,~\cite{Shyam2006}). Indeed,
when we compare ELBM with LBGK using either entopic or standard
polynomial quasiequilibria, there appears to be some gain in
employing entropic quasiequilibria (Fig.~\ref{poly}). We observe
that the post-shock region for the LBGK simulations is more
oscillatory when polynomial quasiequilibria are used. In
Fig.~\ref{poly} we have also included a panel with the simulation
resulting from a much higher viscosity ($\nu=3.3333\times 10^{-2}$).
Here, we observe no appreciable differences in the results of LBGK
and ELBM.

\subsubsection{Nonequilibrium entropy limiters.}

Now, we would like to demonstrate just a representative sample of
the many possibilities of limiters suggested in Sec.~\ref{sec2}. In
each case the limiter is implemented by a post-processing routine
immediately following the collision step (either LBGK~\eqref{lbgkLB}
or ELBM~\eqref{elbmLB}). Here, we will only consider LBGK collisions
and entropic quasiequilibria~\eqref{entropiceq1d}.

The post-processing step adjusts $f$ by the update formula:
\begin{equation*}
    f \mapsto f^*+\phi(\Delta S)(f-f^*),
\end{equation*}
where $\Delta S$ is defined by~\eqref{locaNEQentropy} and $\phi$ is
a limiter function.

For the Ehrenfests' regularisation one would choose
\begin{align*}
\phi(\Delta S)(x) = \left\{
    \begin{aligned}
        &1,\qquad  \Delta S(x) \leq \delta, \\
        &0,\qquad \text{otherwise,}
    \end{aligned}\right.
\end{align*}
where $\delta$ is a pre-specified threshold value. Furthermore, it
is pertinent to select just $k$ sites with highest $\Delta
S>\delta$. This limiter has been previously applied to the shock
tube problem in~\cite{Rob_preprint,Rob,BGJ} and we will not
reproduce those results here.

\begin{figure}
\begin{centering}
\includegraphics[width=13cm]{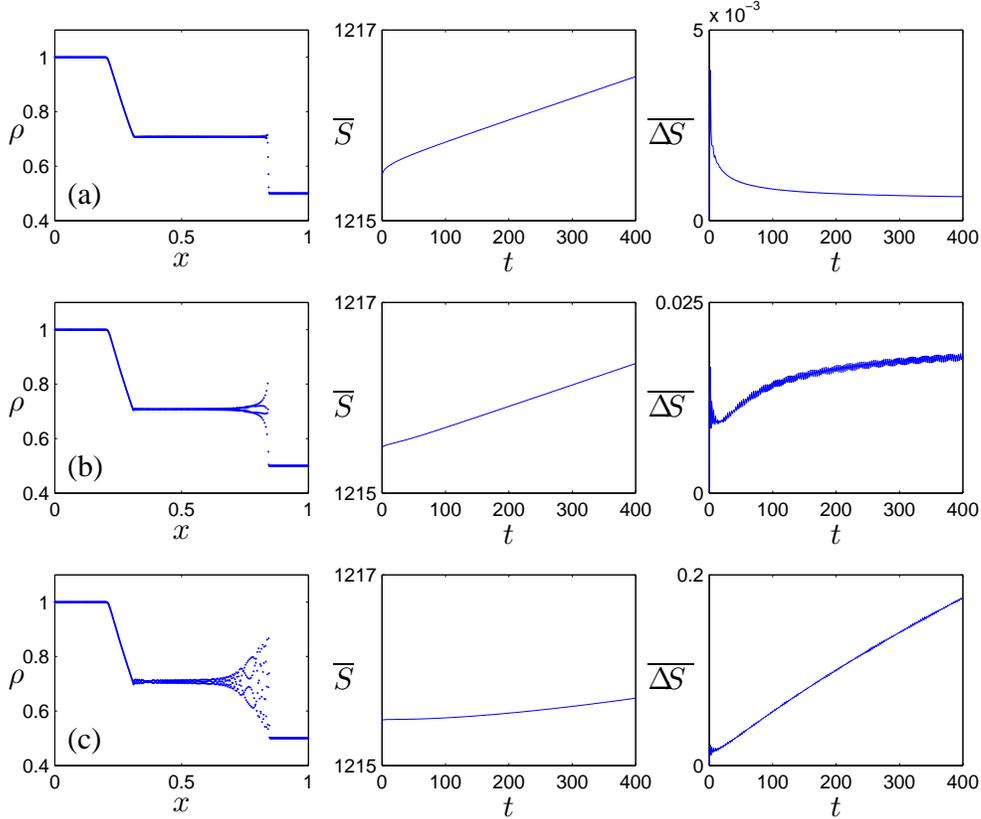}
\caption{Density and profile of the 1:2 athermal shock tube
simulation with $\nu=10^{-9}$ after $400$ time steps using
LBGK~\eqref{lbgkLB} and the smooth limiter~\eqref{smooth1} with
$k=1/2$, $\alpha=\delta/({\rm E} (\Delta S)^k)$ and (a)
$\delta=0.1$; (b) $\delta=0.01$ and (c)
 $\delta=0.001$. Total entropy and nonequilibrium entropy time
histories for each parameter set $\{k,\alpha(\delta)\}$ are
displayed in the adjacent panels.\label{smooth1_fig}}
\end{centering}
\end{figure}

Instead, our first example will be the following smooth limiter:
\begin{equation}\label{smooth1}
    \phi(\Delta S) =  \frac{1}{1+\alpha \Delta S^k}.
\end{equation}
For this limiter, we will fix $k=1/2$ (so that the limiter is double
monotonic in quadratic approximation to entropy) and compare the
density profiles for $\alpha=\delta/({\rm E} (\Delta S)^k)$,
$\delta=0.1,0.01,0.001$. We have also ensured an ensemble-dependent
limiter because of the dependence of $\alpha$ on the average ${\rm
E} (\Delta S)$. As with Fig.~\ref{lbgk_elbm}, we accompany each
panel with the total entropy and nonequilibrium entropy histories.
Note the different scales for nonequilibrium entropy. Note also that
entropy (necessarily) now grows due to the additional dissipation.

\begin{figure}
\begin{centering}
\includegraphics[width=13.0cm]{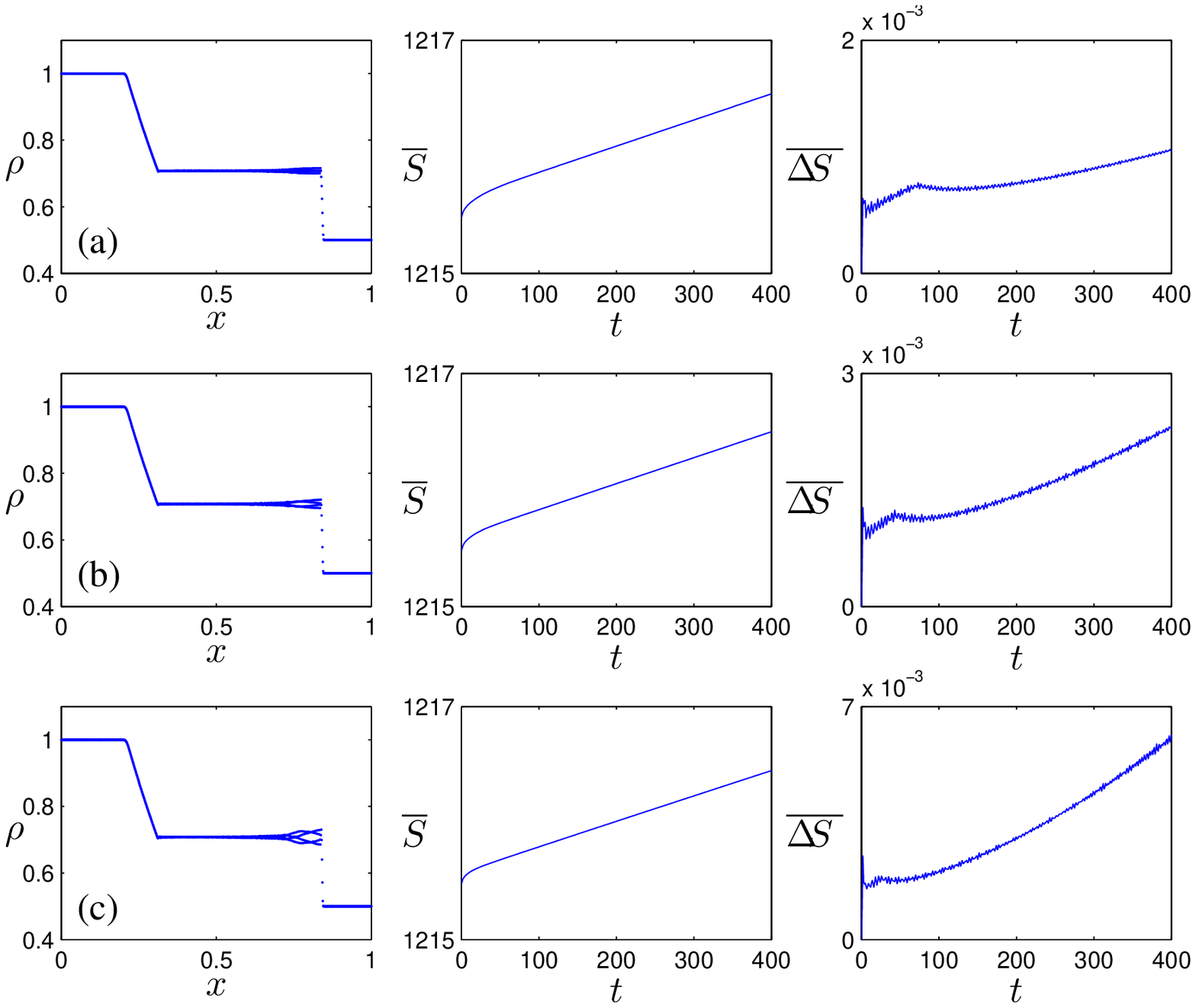}
\caption{Density and profile of the 1:2 athermal shock tube
simulation with $\nu=10^{-9}$ after $400$ time steps using
LBGK~\eqref{lbgkLB} and the threshold limiter~\eqref{EntThresh} with
(a) $\Delta S_t = 5 {\rm E} (\Delta S)$; (b) $\Delta S_t = 10 {\rm
E} (\Delta S)$ and (c) $\Delta S_t = 20 {\rm E} (\Delta S)$. Total
entropy and nonequilibrium entropy time histories for each threshold
$\Delta S_t$ are displayed in the adjacent panels.\label{thres_fig}}
\end{centering}
\end{figure}

Our next example (Fig.~\ref{thres_fig}) considers the threshold
filter~\eqref{EntThresh}. In this example we choose the estimates
$\Delta S_0 =  5{\rm E}(\Delta S), 10{\rm E}(\Delta S), 20{\rm
E}(\Delta S)$ and fix the tolerance $\delta=0$ so that the influence
of the threshold alone can be studied. Only entropic adjustments are
accepted in the limiter: $\Delta S_t \leq \Delta S$. As the
threshold increases, nonequilibrium entropy grows faster and
spurious begin to appear.

\begin{figure}
\begin{centering}
\includegraphics[width=13.0cm]{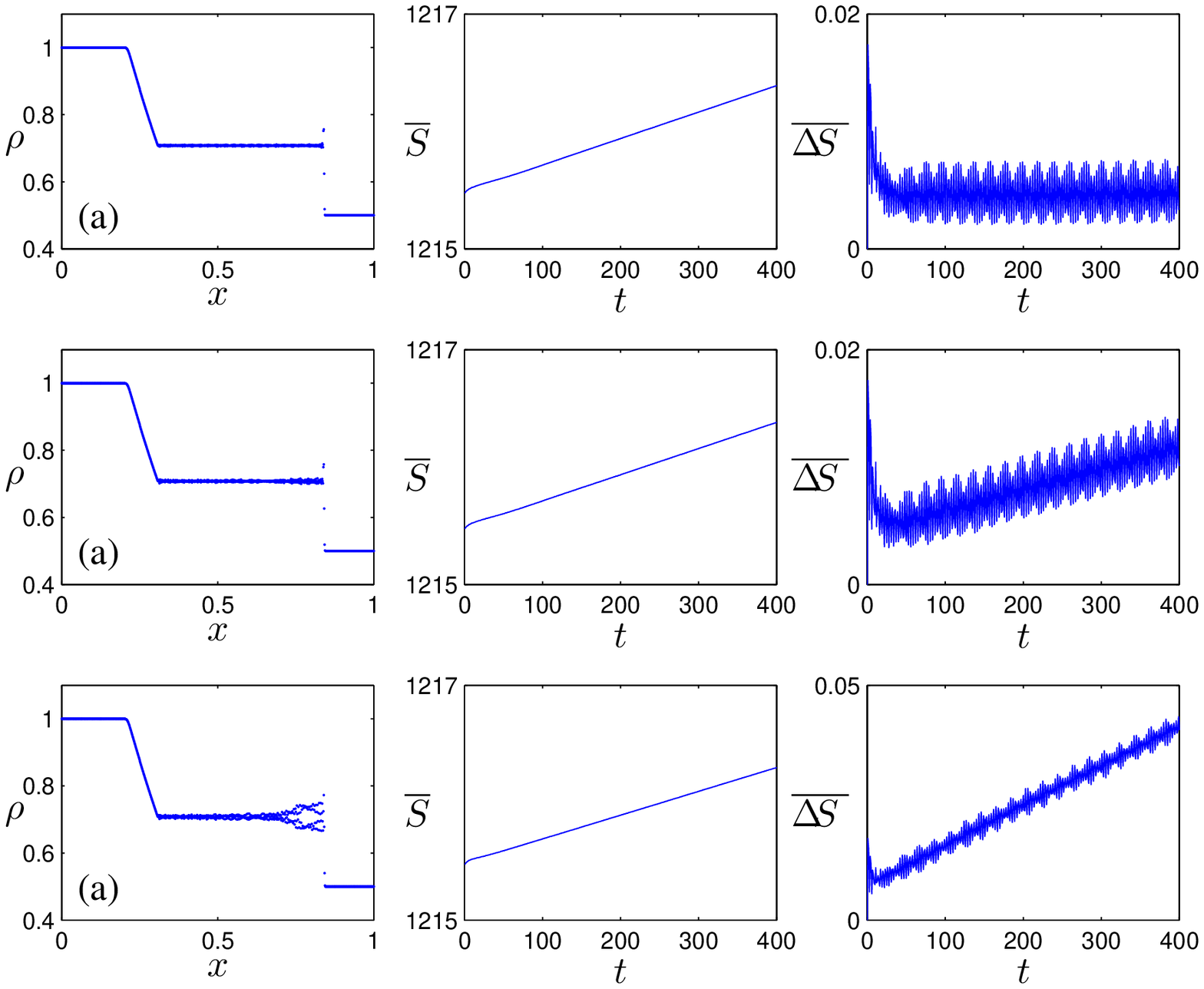}
\caption{Density and profile of the 1:2 athermal shock tube
simulation with $\nu=10^{-9}$ after $400$ time steps using
LBGK~\eqref{lbgkLB} and the minimal median limiter with (a) $\delta
= 10^{-5}$; (b) $\delta =10^{-4}$ and (c) $\delta = 10^{-3}$. Total
entropy and nonequilibrium entropy time histories for each tolerance
$\delta$ are displayed in the adjacent panels.\label{med_fig}}
\end{centering}
\end{figure}

Finally, we test the median filter (Fig.~\ref{med_fig}). We choose a
minimal filter so that only the nearest neighbours are considered.
As with the threshold filter, we introduce a tolerance $\delta$ and
we try the values $\delta=10^{-3},10^{-4},10^{-5}$. Only entropic
adjustments are accepted in the limiter: $\Delta S_{\rm med} \leq
\Delta S$.

We have seen that each of the examples we have considered
(Fig.~\ref{smooth1_fig}, Fig.~\ref{thres_fig} and
Fig.~\ref{med_fig}) is capable of subduing spurious post-shock
oscillations compared with LBGK (or ELBM) on this problem (cf.
Fig.~\ref{lbgk_elbm}). Of course, by limiting nonequilibrium entropy
the result is necessarily an increase in entropy.

From our experiences our recommendation is that the median filter is
the superior choice  amongst all the limiters suggested in
Sec.~\ref{sec2}. The action of the median filter is found to be both
extremely gentle and, at the same time, very effective.

\subsection{Lid-driven cavity}

Our second numerical example is the classical 2D lid-driven cavity
flow. A square cavity of side length $L$ is filled with fluid with
kinematic viscosity $\nu$ (initially at rest) and driven by the
cavity lid moving at a constant velocity $(u_0,0)$ (from left to
right in our geometry).

We will simulate the flow on a $100\times 100$ grid using LBGK
regularised with the median filter limiter. Unless otherwise stated,
we use entropic quasiequilibria~\eqref{maxwellian}. The
implementation of the filter is as follows: the filter is not
applied to boundary nodes; for nodes which immediately neighbour the
boundary the stencil consists of the $3$ nearest neighbours
(including itself) closest to the boundary; for all other nodes the
minimal stencil of $9$ nearest neighbours is used.

We have purposefully selected such a coarse grid simulation because
it is readily found that, on this problem, unregularised LGBK fails
(blows-up) for all but the most modest Reynolds numbers
$\mathrm{Re}:=Lu_0/\nu$.

\subsubsection{Steady-state vortex centres}

For modest Reynolds number the system settles to a steady state in
which the dominant features are a primary central rotating vortex,
with several counter-rotating secondary vortices located in the
bottom-left, bottom-right (and possibly top-left) corners.

Steady state has been extensively investigated in the literature.
The study of Hou et al~\cite{chen95} simulates the flow over a range
of Reynolds numbers using unregularised LBGK on a $256\times 256$
grid. Primary and secondary vortex centre data is provided. We
compare this same statistic for the present median filtered coarse
grid simulation. We will employ the same convergence criteria used
in~\cite{chen95}. Namely, we deem that steady state has been reached
by ensuring that the difference between the maximum value of the
stream function for successive $10,000$ time steps is less that
$10^{-5}$. The stream function, which is not a primary variable in
the LBM simulation, is obtained from the velocity data by
integration using Simpson's rule. Vortex centres are characterised
as local extrema of the stream function.

We compare our results with the LBGK simulations in~\cite{chen95}
and~\cite{Tosi06}. To align ourselves with these studies we specify
the following boundary condition: lid profile is constant; remaining
cavity walls are subject to the ``bounce-back''
condition~\cite{succi01}. In our simulations, the initial uniform
fluid density profile is $\rho=2.7$ and the velocity of the lid is
$u_0=1/10$ (in lattice units).

Collected in Table~\ref{vort}, for $\mathrm{Re}=2000,5000$ and
$7500$, are the coordinates of the primary and secondary vortex
centres using (a) unregularised LBGK; (b) LBGK with median filter
limiter ($\delta=10^{-3}$); (c) LBGK with median filter limiter
($\delta=10^{-4}$), all with non-entropic polynomial
quasiequilibria~\eqref{polyeq2d}. Lines (d), (e) and (f) are the
same but with entropic quasiequilibria~\eqref{maxwellian}. The
remaining lines of Table~\ref{vort} are as follows: (g) literature
data~\cite{chen95} (unregularised LBGK on a $256\times 256$ grid);
(h) literature data~\cite{Tosi06} (positivity rule); (i) literature
data~\cite{Tosi06} (ELBM). With the exception of (g), all simulation
are conducted on a $100\times 100$ grid. The top-left vortex does
not appear at $\mathrm{Re}=2000$ and no data was provided for it
in~\cite{Tosi06} at $\mathrm{Re}=5000$. The unregularised LBGK
$\mathrm{Re}=7500$ simulation blows-up in finite time and the
simulation becomes meaningless. The $y$-coordinate of the two
lower-vortices at $\mathrm{Re}=5000$ in (i) appear anomalously small
and were not reproduced by our experiments with the positivity rule
(not shown).

We have conducted two runs of the experiment with the median filter
parameter $\delta = 10^{-3}$ and $\delta = 10^{-4}$. Despite the
increased number of realisations the vortex centre locations remain
effectively unchanged and we detect no significant variation between
the two runs. This demonstrates the gentle nature of the median
filter. At Reynolds $\mathrm{Re}=2000$ the median filter has no
effect at all on the vortex centres compared with LBGK.

We find no significant differences between the experiments with
entropic and non-entropic polynomial quasiequilibria in this test.

The coordinates of the primary vortex centre for unregularised LBGK
at $\mathrm{Re}=5000$ are already quite inaccurate as LBGK begins to
lose stability. Stability is lost entirely at some critical Reynolds
number $5000< \mathrm{Re} \leq 7500$ and the simulation blows-up.

Furthermore, we have agreement (within grid resolution) with the
data given in~\cite{chen95}. Also compiled in Table~\ref{vort} is
the data from the limiter experiments conducted in~\cite{Tosi06}
(although not explicitly discussed in the language of limiters by
the authors of that work). In~\cite{Tosi06} the authors give vortex
centre data for the positivity rule (Fig.~\ref{PosRule}) and for
ELBM (which we interpret as containing a hidden limiter).
In~\cite{Tosi06} the positivity rule is called FIX-UP.

\begin{table}
\caption{Primary and secondary vortex centre coordinates for the
lid-driven cavity flow at
$\mathrm{Re}=2000,5000,7500$.~\label{vort}}
\begin{small}
\begin{center}
\begin{tabular}{lllllllllllll}
\hline
 && \multicolumn{2}{l}{Primary} && \multicolumn{2}{l}{Lower-left} && \multicolumn{2}{l}{Lower-right} && \multicolumn{2}{l}{Top-left}\\
\cline{3-4}\cline{6-7}\cline{9-10}\cline{12-13}
$\mathrm{Re}$ & & $x$&$y$ && $x$&$y$ && $x$&$y$ && $x$&$y$\\
\hline
2000 & (a)  & 0.5253&0.5455 && 0.0909&0.1010 && 0.8384&0.1010 && \multicolumn{2}{l}{Not applicable}\\
2000 & (b)  & 0.5253&0.5455 && 0.0909&0.1010 && 0.8384&0.1010 && \multicolumn{2}{l}{Not applicable}\\
2000 & (c)  & 0.5253&0.5455 && 0.0909&0.1010 && 0.8384&0.1010 && \multicolumn{2}{l}{Not applicable}\\
2000 & (d)  & 0.5253&0.5455 && 0.0909&0.1010 && 0.8384&0.1010 && \multicolumn{2}{l}{Not applicable}\\
2000 & (e)  & 0.5253&0.5455 && 0.0909&0.1010 && 0.8384&0.1010 && \multicolumn{2}{l}{Not applicable}\\
2000 & (f)  & 0.5253&0.5455 && 0.0909&0.1010 && 0.8384&0.1010 && \multicolumn{2}{l}{Not applicable}\\
2000 & (g)  & 0.5255&0.5490 && 0.0902&0.1059 && 0.8471&0.0980 && \multicolumn{2}{l}{Not applicable}\\
2000 & (h)  & 0.5200&0.5450 && 0.0900&0.1000 && 0.8300&0.0950 && \multicolumn{2}{l}{Not applicable}\\
2000 & (i)  & 0.5200&0.5500 && 0.0890&0.1000 && 0.8300&0.1000 && \multicolumn{2}{l}{Not applicable}\\
\hline
5000 & (a)  & 0.5152&0.6061 && 0.0808&0.1313 && 0.7980&0.0707 && 0.0505&0.8990\\
5000 & (b)  & 0.5152&0.5354 && 0.0808&0.1313 && 0.8081&0.0808 && 0.0606&0.8990\\
5000 & (c)  & 0.5152&0.5354 && 0.0808&0.1313 && 0.8081&0.0808 && 0.0707&0.8889\\
5000 & (d)  & 0.5152&0.5960 && 0.0808&0.1313 && 0.8081&0.0808 && 0.0505&0.8990\\
5000 & (e)  & 0.5152&0.5354 && 0.0808&0.1313 && 0.8081&0.0808 && 0.0606&0.8990\\
5000 & (f)  & 0.5152&0.5354 && 0.0808&0.1313 && 0.8081&0.0808 && 0.0707&0.8889\\
5000 & (g)  & 0.5176&0.5373 && 0.0784&0.1373 && 0.8078&0.0745 && 0.0667&0.9059\\
5000 & (h)  & 0.5150&0.5680 && 0.0950&0.0100 && 0.8450&0.0100 && \multicolumn{2}{l}{Not available}\\
5000 & (i)  & 0.5150&0.5400 && 0.0780&0.1350 && 0.8050&0.0750 && \multicolumn{2}{l}{Not available}\\
\hline
7500 & (a)  & ---&--- && ---&--- && ---&--- && ---&---\\
7500 & (b)  & 0.5051&0.5354 && 0.0707&0.1515 && 0.7879&0.0707 && 0.0606&0.8990\\
7500 & (c)  & 0.5051&0.5354 && 0.0707&0.1515 && 0.7879&0.0707 && 0.0707&0.8889\\
7500 & (d)  & ---&--- && ---&--- && ---&--- && ---&---\\
7500 & (e)  & 0.5051&0.5354 && 0.0707&0.1515 && 0.7879&0.0707 && 0.0606&0.8990\\
7500 & (f)  & 0.5051&0.5354 && 0.0707&0.1515 && 0.7879&0.0707 && 0.0707&0.8889\\
7500 & (g)  & 0.5176&0.5333 && 0.0706&0.1529 && 0.7922&0.0667 && 0.0706&0.9098\\
\hline
\end{tabular}
\end{center}
\end{small}
\end{table}

As Reynolds number increases the flow in the cavity is no longer
steady and a more complicated flow pattern emerges. On the way to a
fully developed turbulent flow, the lid-driven cavity flow is known
to undergo a series of period doubling Hopf bifurcations. On our
coarse grid, we observe that the coordinates of the primary vortex
centre (maximum of the stream function) is a very robust feature of
the flow, with little change between coordinates (no change in
$y$-coordinates) computed at $\mathrm{Re}=5000$ and
$\mathrm{Re}=7500$ with the median filter. On one hand, because of
this observation it becomes inconclusive whether the median limiter
is adding too much additional dissipation. On the other hand, a more
studious choice of control criteria may indicate that the first
bifurcation has already occurred by $\mathrm{Re}=7500$.

\subsubsection{First Hopf bifurcation}

A survey of available literature reveals that the precise value of
$\mathrm{Re}$ at which the first Hopf bifurcation occurs is somewhat
contentious, with most current studies (all of which are for
incompressible flow) ranging from around
$\mathrm{Re}=7400$--$8500$~\cite{bruneau06,pan00,peng03}. Here, we
do not intend to give a precise value because it is a well observed
grid effect that the critical Reynolds number increases (shifts to
the right) with refinement (see, e.g., Fig.~3 in~\cite{peng03}).
Rather, we will be content to localise the first bifurcation and, in
doing so, demonstrate that limiters are capable of regularising
without effecting fundamental flow features.

To localise the first bifurcation we take the following algorithmic
approach. Entropic quasiequilibria are in use. The initial uniform
fluid density profile is $\rho=1.0$ and the velocity of the lid is
$u_0=1/10$ (in lattice units). We record the unsteady velocity data
at a single control point with coordinates $(L/16,13L/16)$ and run
the simulation for $5000$ non-dimensionless time units ($5000L/u_0$
time steps). Let us denote the final 1\% of this signal by
$(u_\mathrm{sig},v_\mathrm{sig})$. We then compute the \emph{energy}
$E_u$ ($\ell_2$-norm normalised by non-dimensional signal duration)
of the deviation of $u_\mathrm{sig}$ from its mean:
\begin{equation}\label{energy}
    E_u :=   \biggl\| \sqrt{\frac{L}{u_0 |u_{\mathrm{sig}}|}} (u_{\mathrm{sig}}- \overline{u_{\mathrm{sig}}}) \biggr\|_{\ell_2},
\end{equation}
where $|u_{\mathrm{sig}}|$ and $\overline{u_{\mathrm{sig}}}$ denote
the length and mean of $u_{\mathrm{sig}}$, respectively. We choose
this robust statistic instead of attempting to measure signal
amplitude because of numerical noise in the LBM simulation. The
source of noise in LBM is attributed to the existence of an
inherently unavoidable neutral stability direction in the numerical
scheme (see, e.g.,~\cite{BGJ}).

We opt not to employ the ``bounce-back'' boundary condition used in
the previous steady state study. Instead we will use the diffusive
Maxwell boundary condition (see, e.g.,~\cite{cercignani75}), which
was first applied to LBM in~\cite{ansumali02}. The essence of the
condition is that populations reaching a boundary are reflected,
proportional to equilibrium, such that mass-balance (in the bulk)
and detail-balance are achieved. The boundary condition coincides
with ``bounce-back'' in each corner of the cavity.

To illustrate, immediately following the advection of populations
consider the situation of a wall, aligned with the lattice, moving
with velocity $u_\mathrm{wall}$ and with outward pointing normal to
the wall in the negative $y$-direction (this is the situation on the
lid of the cavity with $u_\mathrm{wall}=u_0$). The implementation of
the diffusive Maxwell boundary condition at a boundary site $(x,y)$
on this wall consists of the update
\begin{equation*}
    f_i(x,y,t+1) = \gamma f^{*}_i(u_\mathrm{wall}),\qquad i=4,7,8,
\end{equation*}
with
\begin{equation*}
  \gamma = \frac{f_2(x,y,t)+f_5(x,y,t)+f_6(x,y,t)}{f^{*}_4(u_\mathrm{wall})+
  f^{*}_7(u_\mathrm{wall})+f^{*}_8(u_\mathrm{wall})}.
\end{equation*}
Observe that, because density is a linear factor of the
quasiequilibria~\eqref{maxwellian}, the density of the wall is
inconsequential in the boundary condition and can therefore be taken
as unity for convenience. As is usual, only those populations
pointing in to the fluid at a boundary site are updated. Boundary
sites do not undergo the collisional step that the bulk of the sites
are subjected to.

We prefer the diffusive boundary condition over the often
preferred ``bounce-back'' boundary condition with constant lid
profile. This is because we have experienced difficulty in
separating the aforementioned numerical noise from the genuine
signal at a single control point using ``bounce-back''. We remark
that the diffusive boundary condition does not prevent
unregularised LBGK from failing at some critical Reynolds number
$\mathrm{Re}>5000$.

\begin{figure}
\begin{centering}
\includegraphics[width=12.0cm]{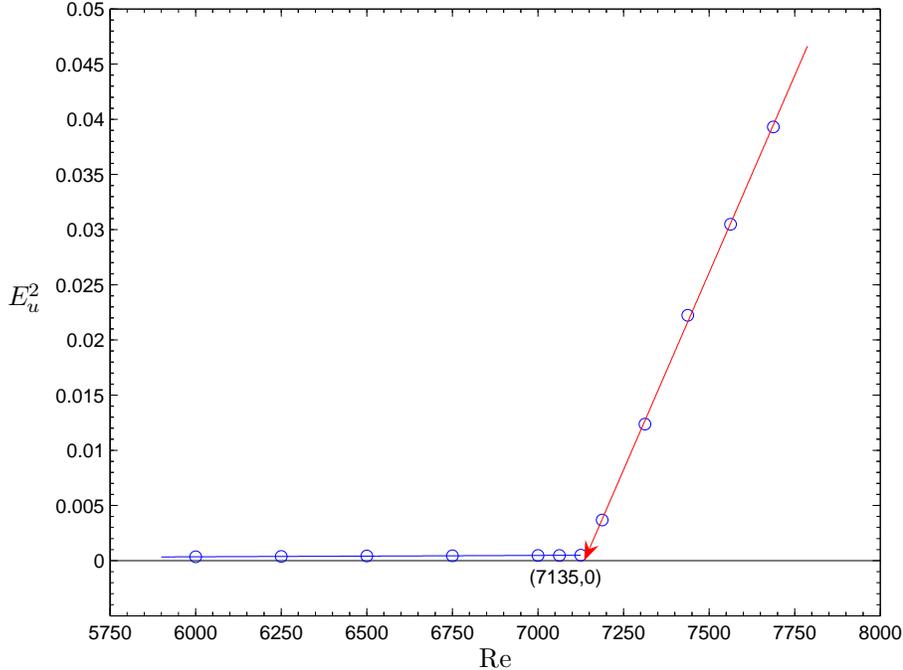}
\caption{Plot of energy squared, $E_u^2$~\eqref{energy}, as a
function of Reynolds number, $\mathrm{Re}$, using LBGK regularised
with the median filter limiter with $\delta = 10^{-3}$ on a $100
\times 100$ grid. Straight lines are lines of best fit. The
intersection of the sloping line with the $x$-axis occurs close to
$\mathrm{Re}=7135$. \label{ERe}}
\end{centering}
\end{figure}

\begin{figure}
\begin{centering}
\includegraphics[width=12.0cm]{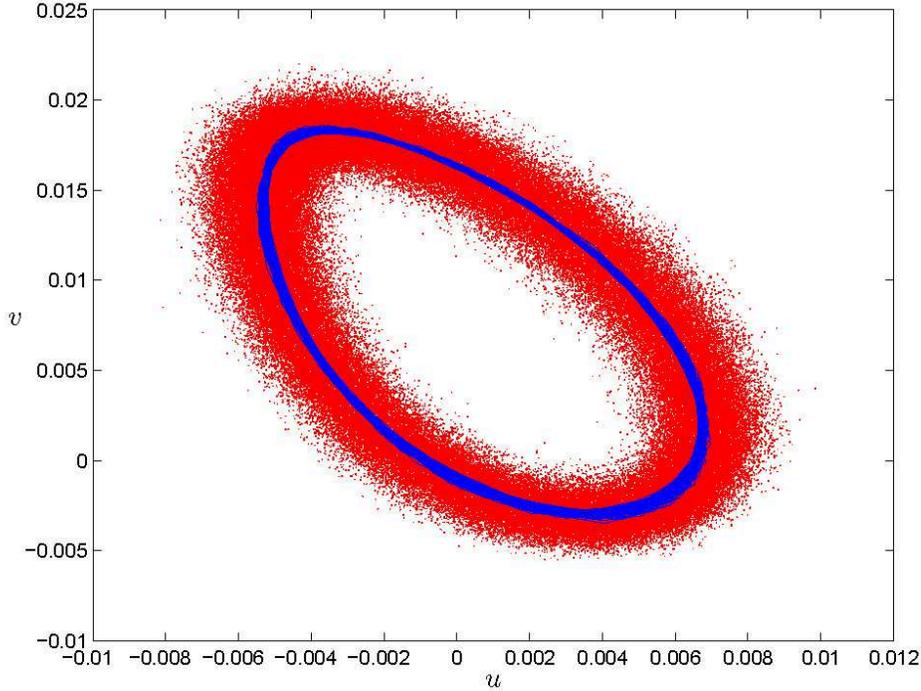}
\caption{Velocity components as a function of time for the signal
$(u_\mathrm{sig},v_\mathrm{sig})$ at the monitoring point
$(L/16,13L/16)$ using LBGK regularised with the median filter
limiter with $\delta = 10^{-3}$ on a $100 \times 100$ grid
($\mathrm{Re}=7375$). Dots represent simulation results and the
solid line is a $100$ step time average of the signal.
\label{phase}}
\end{centering}
\end{figure}

\begin{figure}
\begin{centering}
\includegraphics[width=12.0cm]{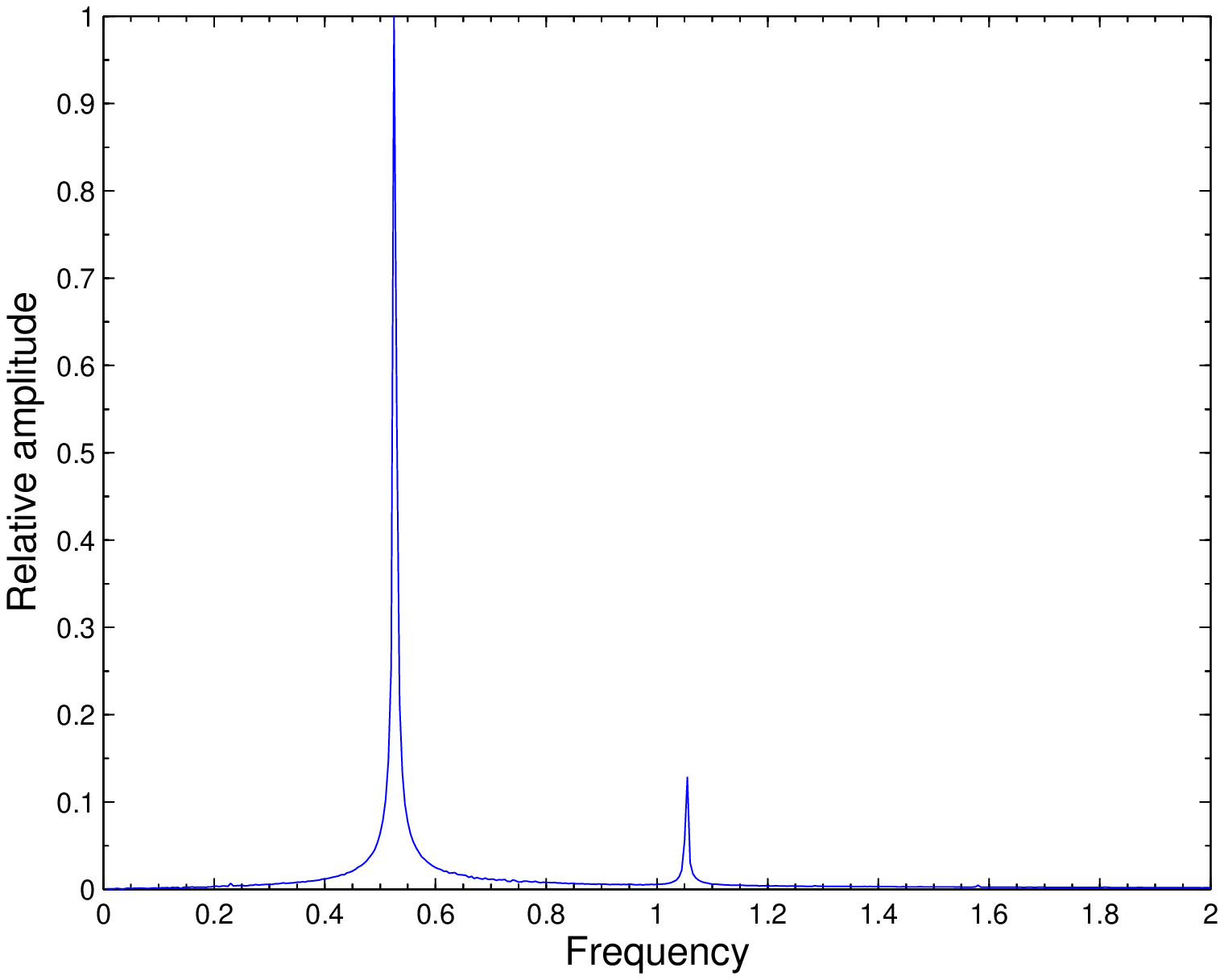}
\caption{Relative amplitude spectrum for the signal
$u_\mathrm{sig}$ at the monitoring point $(L/16,13L/16)$ using
LBGK regularised with the median filter limiter with $\delta =
10^{-3}$ on a $100 \times 100$ grid ($\mathrm{Re}=7375$). We
measure a dominant frequency of $\omega=0.525$.\label{spec}}
\end{centering}
\end{figure}

Now, we conduct an experiment and record~\eqref{energy} over a range
of Reynolds numbers. In each case the median filter limiter is
employed with parameter $\delta=10^{-3}$. Since the transition
between steady and periodic flow in the lid-driven cavity is known
to belong to the class of standard Hopf bifurcations we are assured
that $E_u^2\propto \mathrm{Re}$~\cite{ghadder86}. Fitting a line of
best fit to the resulting data localises the first bifurcation in
the lid-driven cavity flow to $\mathrm{Re}=7135$ (Fig.~\ref{ERe}).
This value is within the tolerance of $\mathrm{Re}=7402\pm4\%$ given
in~\cite{peng03} for a $100 \times 100$ grid. We also provide a
(time averaged) phase space trajectory and Fourier spectrum for
$\mathrm{Re}=7375$ at the monitoring point (Fig.~\ref{phase} and
Fig.~\ref{spec}) which clearly indicate that the first bifurcation
has been observed.

\section{Conclusions}\label{sec4}

Entropy and thermodynamics are important for stability of the
lattice Boltzmann methods. It is now clear: after almost 10 years of
work since the publication of~\cite{KGSBPRL} proved this statement
(the main reviews are~\cite{boghosian01,karlin06,LB3}). The question
is now: ``how does one utilise, optimally, entropy and thermodynamic
structures in lattice Boltzmann methods?''. In our paper we attempt
to propose a solution (temporary, at least). Our approach is
applicable to both entropic as well as for non-entropic polynomial
quasiequilibria.

We have constructed a system of nonequilibrium entropy limiters for
the lattice Boltzmann methods (LBM):
\begin{itemize}
\item{the positivity rule that provides positivity of
distribution;}
\item{the pointwise entropy limiters based on selection and
correction of most nonequilibrium values;}
\item{filters of nonequilibrium entropy, and the median filter as a filter of choice.}
\end{itemize}

All these limiters exploit physical properties of LBM and allow
control of total additional entropy production. In general, they do
the same work for LBM as flux limiters do for finite differences,
finite volumes and finite elements methods, and come into operation
when sharp gradients are present. For smoothly changing waves, the
limiters do not operate and the spatial derivatives can be
represented by higher order approximations without introducing
non-physical oscillations. But there are some differences  too: for
LBM the main idea behind the construction of nonequilibrium entropy
limiter schemes is to limit a scalar quantity --- the nonequilibrium
entropy --- or to delete the ``salt and pepper" noise from the field
of this quantity. We do not touch the vectors or tensors of spatial
derivatives, as it is for flux limiters.

Standard test examples demonstrate that the developed limiters erase
spurious oscillations without blurring of shocks, and do not affect
smooth solutions. The limiters we have tested do not produce a
noticeable additional dissipation and allow us to reproduce the
first Hopf bifurcation for 2D lid-driven cavity on a coarse $100
\times 100$ grid. At the same time the simplest median filter
deletes the spurious post-shock oscillations for low viscosity.

Perhaps, it is impossible to find one best nonequilibrium entropy
limiter for all problems. It is a special task to construct the
optimal limiters for a specific classes of problems.

\section*{Acknowledgments}

Discussion of the preliminary version of this work with S. Succi and
participants of the lattice Boltzmann workshop held on 15th November
2006 in Leicester (UK) was very important. Author A.~N.~Gorban is
grateful to S.~K.~Godunov for the course of numerical methods given
many years ago at Novosibirsk University. This work is supported by
Engineering and Physical Sciences Research Council (EPSRC) grant
number GR/S95572/01.


\begin{thebibliography}{44}

\bibitem{ansumali02}
S.~Ansumali, and I.~V.~Karlin.
\newblock Kinetic boundary conditions in the lattice Boltzmann method.
\newblock { Phys. Rev. E} {\bf 66}, 026311 2002.

\bibitem{ansumali03}
S.~Ansumali S, I.~V.~Karlin, H.~C.~Ottinger.
\newblock Minimal entropic kinetic models for hydrodynamics
\newblock {\em Europhys. Let.} 63 (6): 798-804. 2003

\bibitem{Benzi}
R.~Benzi, S.~Succi, and M.~Vergassola.
\newblock The lattice {B}oltzmann-equation - theory and applications.
\newblock {\em Physics Reports}, 222(3):145--197, 1992.

\bibitem{Bruce}
B.~M.~Boghosian, P.~J.~Love, and J.~Yepez.
\newblock Entropic lattice {B}oltzmann model for {B}urgers equation.
\newblock {\em Phil. Trans. Roy. Soc. A}, 362:1691--1702, 2004.

\bibitem{boghosian01}
B.~M.~Boghosian, J.~Yepez, P.~V.~Coveney, and A.~J.~Wager.
\newblock Entropic lattice {B}oltzmann methods.
\newblock {\em R. Soc. Lond. Proc. Ser. A Math. Phys. Eng. Sci.},
  457(2007):717--766, 2001.

\bibitem{Rob_preprint}
R.~A.~Brownlee, A.~N.~Gorban, and J.~Levesley.
\newblock Stabilisation of the lattice-{B}oltzmann method using the
  {E}hrenfests' coarse-graining.
\newblock {\em cond-mat/0605359}, 2006.

\bibitem{Rob}
R.~A. Brownlee, A.~N.~Gorban, and J.~Levesley.
\newblock Stabilisation of the lattice-{B}oltzmann method using the
  {E}hrenfests' coarse-graining.
\newblock {\em Phys. Rev. E}, 74:037703, 2006.

\bibitem{BGJ}R.~A.~Brownlee, A.N.~Gorban, and J.~Levesley.
Stability and stabilization of the lattice Boltzmann method, Phys.
Rev. E, to appear. {\em cond-mat/0611444}, 2006.

\bibitem{bruneau06}
C.-H.~Bruneau, and M.~Saad.
\newblock The 2D lid-driven cavity problem revisited.
\newblock {\em Comput. Fluids}, 35:326--348, 2006.

\bibitem{ChatOsher1983}S.~R.~Chatkravathy, and S.~Osher.  High resolution
applications of the Osher upwind scheme for the Euler equations,
AIAA Paper 83-1943, Proc. AIAA 6th Comutational Fluid Dynamics
Conference, (1983), 363--373.

\bibitem{cercignani75}
C.~Cercignani.
\newblock {\em Theory and Application of the Boltzmann Equation}.
\newblock Scottish Academic Press, Edinburgh, 1975.

\bibitem{LB1}
S.~Chen and G.~D.~Doolen.
\newblock Lattice Boltzmann method for fluid flows.
\newblock {\em Annu. Rev. Fluid. Mech.}, 30:329--364, 1998.

\bibitem{Shyam2006}
S.~S.~Chikatamarla and I.~V.~Karlin. Entropy and Galilean Invariance
of Lattice Boltzmann Theories. {\em Phys. Rev. Lett.} 97, 190601
(2006)

\bibitem{Raz}A.~J.~Chorin, O.~H.~Hald, R.~Kupferman.
Optimal prediction with memory,  {\it Physica D} 166 (2002),
239--257.

\bibitem{ehrenfest11}
P.~Ehrenfest and T.~Ehrenfest.
\newblock {\em The conceptual foundations of the statistical approach in
  mechanics}.
\newblock Dover Publications Inc., New York, 1990.

\bibitem{ghadder86}
N.~K.~Ghaddar, K.~Z.~Korczak, B.~B.~Mikic, and A.~T.~Patera.
\newblock Numerical investigation of incompressible flow in grooved channels. Part 1. Stability and self-sustained
oscillations.
\newblock {\em J. Fluid Mech.}, 163:99--127, 1986.

\bibitem{Godunov}S.~K.~Godunov.  A Difference Scheme for Numerical
Solution of Discontinuous Solution of Hydrodynamic Equations,
Math. Sbornik, 47 (1959), 271-306.

\bibitem{G1}A.~N.~Gorban. Equilibrium encircling. Equations of chemical kinetics and their
thermodynamic analysis, Nauka, Novosibirsk, 1984.

\bibitem{GKOeTPRE2001}
A.~N.~Gorban, I.~V.~Karlin, H.~C.~\"{O}ttinger, and
L.~L.~Tatarinova.
\newblock Ehrenfest's argument extended to a formalism of nonequilibrium
  thermodynamics.
\newblock {\em Phys. Rev. E}, 62:066124, 2001.

\bibitem{gorban06}
A.~N.~Gorban.
\newblock Basic types of coarse-graining.
\newblock In A.~N. Gorban, N.~Kazantzis, I.~G. Kevrekidis, H.-C. \"{O}ttinger,
  and C.~Theodoropoulos, editors, {\em Model Reduction and Coarse-Graining
  Approaches for Multiscale Phenomena}, pages 117--176. Springer,
  Berlin-Heidelberg-New York, 2006.
\newblock cond-mat/0602024.

\bibitem{Kagan} A.~Gorban, B.~Kaganovich, S.~Filippov, A.~Keiko, V.~Shamansky,
I.~Shirkalin, {\em Thermodynamic Equilibria and Extrema: Analysis
of Attainability Regions and Partial Equilibrium}, Springer,
Berlin, Heidelberg, New York,   2006.

\bibitem{Grad} H.~Grad.
On the kinetic theory of rarefied gases, {\it Comm. Pure and Appl.
Math.} {2} 4, (1949), 331--407.

\bibitem{Higuera}
F.~Higuera, S.~Succi, and R.~Benzi.
\newblock Lattice gas -- dynamics with enhanced collisions.
\newblock {\em Europhys. Lett.}, 9:345--349, 1989.

\bibitem{chen95}
S.~Hou, Q.~Zou, S.~Chen, G.~Doolen and A.~C.~Cogley.
\newblock Simulation of cavity flow by the lattice {B}oltzmann method.
\newblock {\em J. Comp. Phys.}, 118:329--347, 1995.

\bibitem{EIT} D.~Jou, J.~Casas-V\'azquez, G.~Lebon.
{\em Extended irreversible thermodynamics}, Springer, Berlin, 1993.

\bibitem{KGSBPRL}
I.~V.~Karlin, A.~N.~Gorban, S.~Succi, and V.~Boffi.
\newblock Maximum entropy principle for lattice kinetic equations.
\newblock {\em Phys. Rev. Lett.}, 81:6--9, 1998.

\bibitem{karlin99}
I.~V.~Karlin, A.~Ferrante, and H.~C.~\"{O}ttinger.
\newblock Perfect entropy functions of the lattice {B}oltzmann method.
\newblock {\em Europhys. Lett.}, 47:182--188, 1999.

\bibitem{karlin06}
I.~V.~Karlin, S.~Ansumali, C.~E.~ Frouzakis, and S.~S.~Chikatamarla.
\newblock Elements of the lattice Boltzmann method I: Linear advection
equation.
\newblock {\em Commun. Comput. Phys.}, 1 (2006), 616--655.

\bibitem{karlin07} I.~V.~Karlin, S.~S.~Chikatamarla and S.~Ansumali.
\newblock Elements of the lattice Boltzmann method II: Kinetics and hydrodynamics in one
dimension.
\newblock {\em Commun. Comput. Phys.}, 2 (2007), 196--238.

\bibitem{Kull}S.~Kullback. Information theory and statistics, Wiley, New York, 1959.

\bibitem{li04} Y.~Li, R.~Shock, R.~Zhang, and H.~Chen.
\newblock Numerical study of flow past an impulsively
started cylinder by the lattice-Boltzmann method.
\newblock {\em J. Fluid Mech.}, 519:273--300, 2004.

\bibitem{pan00}
T.~W.~Pan, and R.~Glowinksi.
\newblock A projection/wave-like equation method for the numerical
simulation of incompressible viscous fluid flow modeled by the
Navier--Stokes equations.
\newblock {\em Comp. Fluid Dyn. J.}, 9:28--42, 2000.

\bibitem{peng03}
Y.-F.~Peng, Y.-H. Shiau, and R.~R.~Hwang.
\newblock Transition in a 2-D lid-driven cavity flow.
\newblock {\em Comput. Fluids}, 32:337--352, 2003.

\bibitem{Pratt}W.~K.~Pratt. Digital Image Processing, Wiley, New York, 1978.

\bibitem{Qian} H.~Qian. Relative entropy: free energy associated with equilibrium
fluctuations and nonequilibrium deviations, Phys. Rev. E. 63 (2001),
042103.

\bibitem{Roe1986}P.~L.~Roe. Characteristic-based schemes for the Euler
equations, {\em Ann. Rev. Fluid Mech.}, 18 (1986), 337-365.

\bibitem{HudongGrad} X.~Shan, X-F.~Yuan, and H.~Chen. Kinetic theory representation of
hydrodynamics: a way beyond the Navier–Stokes equation. J. Fluid
Mech. 550 (2006), 413-–441.

\bibitem{succi01}
S.~Succi.
\newblock {\em The lattice {B}oltzmann equation for fluid dynamics and beyond}.
\newblock Oxford University Press, New York, 2001.

\bibitem{LB3}
S.~Succi, I.~V.~Karlin, and H.~Chen.
\newblock Role of the {H} theorem in lattice {B}oltzmann hydrodynamic
  simulations.
\newblock {\em Rev. Mod. Phys.}, 74:1203--1220, 2002.

\bibitem{Sweby1984}P.~K.~Sweby. High resolution schemes using
flux-limiters for hyperbolic conservation laws. {\em SIAM J. Num.
Anal.}, 21 (1984), 995--1011.

\bibitem{Tosi06}
F.~Tosi, S.~Ubertini, S.~Succi, H.~Chen, and I.V. Karlin.
\newblock Numerical stability of entropic versus positivity-enforcing lattice
  {B}oltzmann schemes.
\newblock {\em Math. Comput. Simulation}, 72:227--231, 2006.

\bibitem{vanLeer1977}B.~Van Leer. Towards the ultimate conservative
difference scheme III. Upstream-centered finite-difference schemes
for ideal compressible flow., J. Comp. Phys., 23 (1977), 263--275.

\bibitem{Wess}P.~Wesseling. Principles of Computational Fluid
Dynamics,  Springer Series in Computational Mathematics
(Springer-Verlag, Berlin, 2001), Vol. 29.

\bibitem{Zeld}Y.~B.~Zeldovich, Proof of the Uniqueness of the Solution of the
Equations of the Law of Mass Action, In: {\it Selected Works of
Yakov Borisovich Zeldovich,} Vol. 1, J.~P.~Ostriker (Ed.), Princeton
University Press, Princeton, USA, 1996, 144--148.

\end{thebibliography}
\end{document}